\documentclass{article}
\usepackage{amssymb}
\usepackage{amsmath}
\usepackage{graphicx}

\input{tcilatex}

\begin{document}

\title{Static spherically symmetric collapsar in the relativistic theory of
gravitation: numerical approach and stability analysis}
\author{Vladimir L. Kalashnikov \\
Photonics Institute, TU Vienna, Gusshausstr. 27/387,\\
A-1040 Vienna, Austria\\
tel: +43-158-8013-8723, fax: +43-158-8013-8799,\\
e-mail:kalashnikov@tuwien.ac.at}
\date{}
\maketitle

\begin{abstract}
Numerical investigation of the static spherically symmetric vacuum solution
of the Logunov equations confirms the analytical results and demonstrates a
strong repulsion at sub-Planckian distance from the Schwarz-schild-like
singularity, which is stable against the small perturbations. We suppose
that the final stage of the stellar collapse in the relativistic theory of
gravitation can entail the gravitational bursts with the gravitons
overproduction due to the strong deceleration in the vicinity of the
Schwarzschild sphere. The rigorous investigation of this topic requires a
quantum-field consideration.
\end{abstract}

\section{Introduction}

The relativistic theory of gravitation (RTG) is built as a classical field
theory in the Minkowski spacetime \cite{logunov1,logunov2,logunov3,logunov4}%
. Since the gravitational field is described by the symmetric second-rank
tensor, it manifests itself as an effective Riemannian spacetime. Thus, the
RTG is the bi-metric theory, which needs the gauge-symmetry violation to be
physically self-consistent. This requires a massive graviton. As a result,
the field equations contain both metrics (i.e. both Riemannian and
Minkowski) and gauge-fixing condition, which eliminates the undesirable
spin-states of graviton. The cosmological consequences of the non-zero
graviton mass impose a limitation on its value, which cannot exceed $%
10^{-66} $ g \cite{gershtein}, but virtually can be incomparably smaller
than this value \cite{kalashnikov1}.

The non-zero mass of graviton causes the repulsion (antigravitation) at
small distances and the deceleration of the cosmological expansion at its
late stage. For the black-hole physics is most important the antigravitation
prohibiting an unlimited collapse of the massive stellar objects. This topic
had been considered earlier in the pure analytical framework (see, for
example, \cite{logunov5}). However, the RTG field equations (Logunov field
equations) don't allow an exact analytical solution even for the static
spherically symmetric field and need some approximations in order to be
solved in a closed form. Moreover, the physical consistency of the obtained
solutions needs their stability at least against the small perturbations.

In this work we consider the numerical solutions of the Logunov field
equations in the vicinity of the Schwarzschild sphere and compare their with
the analytical approximation. This comparison demonstrates a high accuracy
of the latter. Also, we analyze the stability of the analytical metric in
the vicinity of the Schwarzschild sphere. Although this metric provides a
strong repulsion for the radially falling particle, it is stable against the
small perturbations.

\section{Numerical approach to the Schwarzschild-like singularity}

The Logunov equations for the massive gravitational field have the following
form \cite{logunov1,logunov2,logunov3,logunov4}:

\begin{gather}
G_{\nu }^{\mu }-\frac{m^{2}}{2}\left( \delta _{\nu }^{\mu }+g^{\mu \lambda
}\gamma _{\lambda \nu }-\frac{1}{2}\delta _{\nu }^{\mu }g^{\kappa \lambda
}\gamma _{\kappa \lambda }\right) =-\frac{8\pi \varkappa }{c^{4}}T_{\nu
}^{\mu },  \label{eq1} \\
D_{\mu }\hat{g}^{\mu \nu }=0.  \notag
\end{gather}

\noindent Here $g^{\mu \lambda }$\ is the metric tensor of the effective
Riemannian spacetime, $\hat{g}^{\mu \nu }=\sqrt{-g}g^{\mu \nu }$; $\gamma
^{\mu \nu }$\ is the metric tensor of the background Minkowski spacetime, $%
G_{\nu }^{\mu }$\ is the Einstein tensor, $T_{\nu }^{\mu }$\ is the matter
energy-momentum tensor, $D_{\mu }$\ is the covariant derivative in the
Minkowski spacetime; $m^{2}=\left( m_{g}c\diagup \hslash \right) ^{2}$, $%
m_{g}$ is the graviton mass; $\varkappa $ is the Newtonian gravitational
constant.

Spherically symmetric static interval of the effective Riemannian spacetime
can be written in the following form:

\begin{equation}
ds^{2}=U\left( r\right) dt^{2}-V\left( r\right) dr^{2}-W\left( r\right)
^{2}\left( d\theta ^{2}+\sin ^{2}\theta \ d\varphi ^{2}\right) .  \label{eq2}
\end{equation}

Henceforward it is convenient to use the dimensionless quantities and we
shall normalize the geometric units to the Schwarzschild radius $%
r_{s}=2\varkappa M\diagup c^{2}$: $W=x\ r_{s}$ and $r=y\ r_{s}$. Then the
field equations (\ref{eq1}) written in the interval (\ref{eq2}) allow the
system of the master equations ($\epsilon =\frac{1}{2}\left( \frac{%
2\varkappa Mm_{g}}{\hslash c}\right) ^{2}$, $\zeta =\frac{8\pi \varkappa
r_{s}^{2}}{c^{2}}$) \cite{kalashnikov2}:

\begin{gather}
\frac{\partial }{\partial x}\frac{x}{V\left( \frac{\partial y}{\partial x}%
\right) ^{2}}-1+\epsilon \left[ y^{2}-x^{2}+\frac{x^{2}}{2}\left( \frac{1}{V}%
-\frac{1}{U}\right) \right] =-\zeta \rho x^{2},  \notag \\
x\frac{\frac{\partial }{\partial x}\ln \left( xU\right) }{V\left( \frac{%
\partial y}{\partial x}\right) ^{2}}-1+\epsilon \left[ y^{2}-x^{2}+\frac{%
x^{2}}{2}\left( \frac{1}{U}-\frac{1}{V}\right) \right] =\zeta px^{2},
\label{eq3} \\
\frac{\partial }{\partial x}\sqrt{\frac{U}{V}}x^{2}=2y\sqrt{UV}\frac{%
\partial y}{\partial x},  \notag \\
\frac{\partial p}{\partial x}=-\frac{1}{2}\left( p+\rho \right) \frac{\frac{%
\partial U}{\partial x}}{U}.  \notag
\end{gather}

\noindent Here we supposed that $T_{0}^{0}=\rho $, $T_{m}^{m}=-p\diagup
c^{2} $ ($m=1..3$).

Using the new functions we write \cite{logunov5}:

\begin{equation}
U=\frac{1}{x\eta A},~V=\frac{x}{A\left( \frac{dy}{dx}\right) ^{2}}.
\label{eq4}
\end{equation}

\noindent Since $\epsilon $ is an extremely small quantity (see below), we
can suppose

\begin{equation}
x,\ y\ll \frac{1}{\sqrt{2\epsilon }}.  \label{eq5}
\end{equation}

\noindent Then in the vacuum from (\ref{eq3},\ref{eq4},\ref{eq5}) we have

\begin{gather}
A\left( x\right) \frac{d\ln \left[ \eta \left( x\right) \right] }{dx}+2=0,
\label{eq6} \\
\frac{d}{dx}A\left( x\right) =1+\frac{1}{2}\epsilon x^{2}A\left( x\right) %
\left[ x\eta \left( x\right) -\frac{1}{x}\right] .  \notag
\end{gather}

Several approximations based on the assumption $x\approx x_{1}$ (where $%
x_{1}>1$\ is the Schwarzschild-like singularity point) allow the analytical
expression for the interval in the vicinity of the Schwarzschild sphere \cite%
{logunov5,kalashnikov3}:

\begin{equation}
ds^{2}=\frac{\epsilon }{2}dt^{2}-\frac{dy^{2}}{2\left( 1-\frac{x_{1}}{y}%
\right) }-y^{2}\left( d\theta ^{2}+\sin ^{2}\theta \ d\varphi ^{2}\right) .
\label{eq7}
\end{equation}

\noindent The obvious differences from the Schwarzschild metric result in
the main physical consequence: the radially falling particle is decelerated
in the vicinity of $x_{1}$\ and cannot reach the singularity. This statement
is in accord with the causality principle in the RTG, which requires $%
x_{0}>x_{1}$ ($x_{0}$ is the position of the collapsar surface).

To avoid the approximations required for the analytical expression (\ref{eq7}%
), system (\ref{eq6}) was considered numerically. Since the upper threshold
for the graviton mass is $1.6\times 10^{-66}$ g \cite{gershtein}, $\epsilon $%
\ has an extremely small value for the stellar objects (e.g. $\epsilon
\simeq 10^{-44}$ for $M=10M_{\odot }$). This allows the assumption $\frac{dy%
}{dx}=1$ and Eqs. (\ref{eq4},\ref{eq6}) result in:

\begin{equation}
2\eta \left( x\right) \frac{d^{2}\eta \left( x\right) }{dx^{2}}-3\left(
\frac{d\eta \left( x\right) }{dx}\right) ^{2}-\epsilon x\eta \left( x\right)
\left( 1-x^{2}\eta \left( x\right) \right) \frac{d\eta \left( x\right) }{dx}%
=0.  \label{eq8}
\end{equation}

\noindent \qquad Far off $x_{1}$ or for $\epsilon =0$ Eq. (\ref{eq8}) leads
to the Schwarzschild metric. The behavior of $U$ and $V$ in the vicinity of $%
x_{1}$ for nonzero $\epsilon $ is shown in Fig. 1 (dashed curves present the
Schwarzschild solution). The point in Fig. 1, \textit{a} and the dotted
curve in Fig. 1, \textit{b} correspond to the analytical interval (\ref{eq7}%
) but for the numerically obtained value of $x_{1}$.

\begin{figure}[tbp]
\centering\includegraphics[width=12cm]{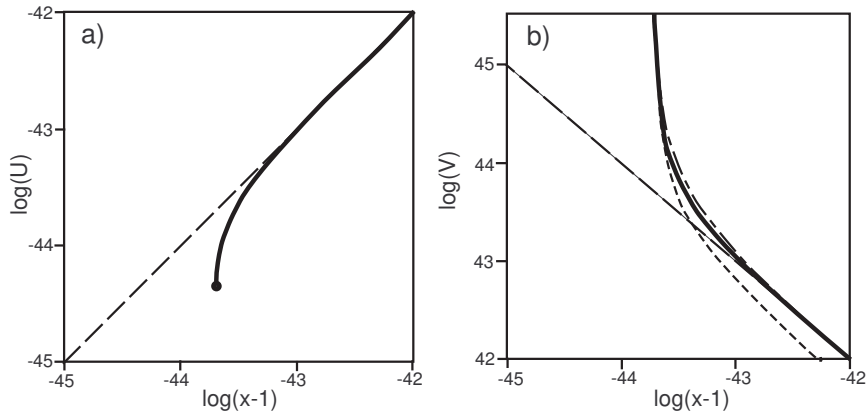}
\caption{Dependence of the dimensionless metric functions on the
dimensionless distance from the Schwarzschild sphere. Solid -- numerical
simulation, dotted -- metric (\protect\ref{eq7}), dashed -- Schwarzschild
metric with singularity at $r_s$, dashed-dotted -- Schwarzschild metric with
singularity at $x_1$. Point in (\textit{a}) corresponds to analytical metric
(\protect\ref{eq7}).}
\label{}
\end{figure}

\begin{figure}[tbp]
\centering\includegraphics[width=7.5cm]{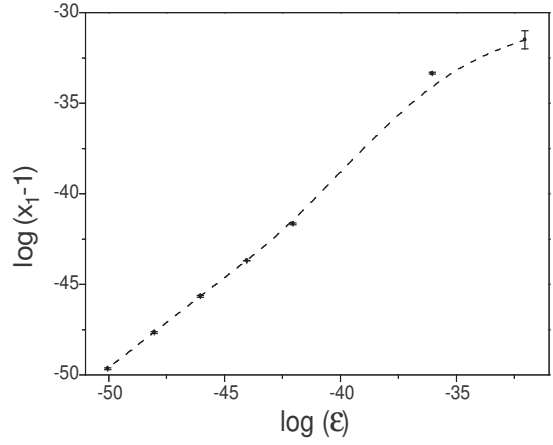}
\caption{Distance of the singularity point from the Schwarzschild sphere
depending on the dimensionless mass parameter $\protect\epsilon$.}
\label{}
\end{figure}

\begin{figure}[tbp]
\centering\includegraphics[width=8cm]{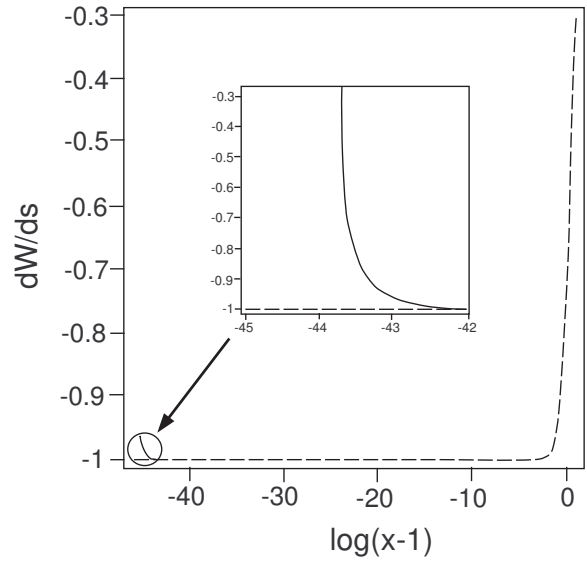} \caption{Velocity
of the falling particle depending on the distance from the
Schwarzschild sphere. Solid -- numerical result from the Logunov
equations, dashed -- result from the Schwarzschild metric.}
\label{}
\end{figure}

Firstly, we have to recognize a good accuracy of the analytical estimation
in the vicinity of $x_{1}$. Note also, that the Schwarzschild solution for $V
$\ (dash-dotted curve in Fig. 1, \textit{b}) is quite accurate if $%
r_{s}=x_{1}$.\ Secondly, the differences between Schwarzschild and Logunov
metrics appear at the extremely small distance from $r_{s}$ (sub-Planckian
for the stellar objects). It is clear that the latter results from the small
value of $m_{g}$ and requires the quantum consideration of the Schwarzschild
sphere vicinity. The dependance of $x_{1}$ on $\epsilon $ is shown in Fig.
2. One can see that the deviation of $x_{1}$ from $r_{s}$ is very small even
for the extra-massive collapsars.

The approach to $x_{1}$ of the free-falling particle is described in the
vicinity of the Schwarzschild sphere by the approximate expression \cite%
{logunov5}:

\begin{equation}
\frac{dW\left( s\right) }{ds}=-2\sqrt{\frac{x-x_{1}}{\epsilon x_{1}^{3}}},
\label{eq9}
\end{equation}

\noindent which demonstrates a strong deceleration due to antigravitation
induced by the massive graviton. Fig. 3 shows the numerical result obtained
from Eqs. (\ref{eq4},\ref{eq8}) and expression

\begin{equation}
\frac{dW\left( s\right) }{ds}=-\sqrt{\frac{1-U}{UV^{\ast }}}  \label{eq10}
\end{equation}

\noindent for the modified interval

\begin{equation}
ds^{2}=U\left( W\right) dt^{2}-V^{\ast }\left( W\right) dW^{2}-W^{2}\left(
d\theta ^{2}+\sin ^{2}\theta \ d\varphi ^{2}\right) .  \label{eq11}
\end{equation}

One can see, that the velocity of the falling particle (for the remote
observer) approaches to the velocity of light ($c=1$ in our normalization).
However, in immediate proximity to $x_{1}$ the particle is decelerated by
the short-range antigravitation and its velocity decreases up to zero (i.e.
the Schwarzschild sphere is physically unreachable). Surely, as this
deceleration takes a place at the extremely small distance, we need a
quantum consideration. Nevertheless, one can suppose that the final stage of
collapse can entail the over-production of the gravitons (some kind of the
gravitational burst). Note also, that in spite of $x_{1}-1\ll 1$ the time of
the light propagation from $r=100r_{s}$ to the point, where the deviation
from the Schwarzschild metric begins, is $\approx 0.02$ s for $M=$ $%
10M_{\odot }$. Thus, the processes induced by the massive graviton in the
vicinity of the Schwarzschild sphere are not hidden from us by the
cosmological-scale time and we face the challenge of the stability of the
observable massive collapsar-like objects \cite{bh}.

\section{Stability of the spherically symmetric metric}

Although the problem of the collapsar stability can not be reduced to the
stability of the metric against the small perturbation, the latter is its
necessary condition. We shall base our analysis on the classical method of
Ref. \cite{chandra}. (\ref{eq7}) is chosen as the perturbed metric.

The general perturbed metric has a form:

\noindent
\begin{gather}
g_{\mu \nu }=  \label{eq12} \\
\begin{bmatrix}
U\left( r\right) , & 0, & 0, &
\begin{array}{c}
\begin{array}{c}
\delta \ \omega \left( t,r,\theta \right) \times \\
W\left( r\right) ^{2}\sin ^{2}\theta ,%
\end{array}
\\
\end{array}
\\
0, & -V\left( r\right) , & 0, &
\begin{array}{c}
\begin{array}{c}
\delta \ q_{2}\left( t,r,\theta \right) \times \\
W\left( r\right) ^{2}\sin ^{2}\theta ,%
\end{array}
\\
\end{array}
\\
0, & 0, & -W\left( r\right) ^{2}, &
\begin{array}{c}
\begin{array}{c}
\delta \ q_{3}\left( t,r,\theta \right) \times \\
W\left( r\right) ^{2}\sin ^{2}\theta ,%
\end{array}
\\
\end{array}
\\
\begin{array}{c}
\delta \ \omega \left( t,r,\theta \right) \times \\
W\left( r\right) ^{2}\sin ^{2}\theta%
\end{array}%
, &
\begin{array}{c}
\delta \ q_{2}\left( t,r,\theta \right) \times \\
W\left( r\right) ^{2}\sin ^{2}\theta%
\end{array}%
, &
\begin{array}{c}
\delta \ q_{3}\left( t,r,\theta \right) \times \\
W\left( r\right) ^{2}\sin ^{2}\theta%
\end{array}%
, & -W\left( r\right) ^{2}\sin ^{2}\theta%
\end{bmatrix}%
,  \notag
\end{gather}

\noindent where $\delta $\ is the small expansion parameter; $\omega \left(
t,r,\theta \right) $, $q_{2}\left( t,r,\theta \right) $ and $q_{3}\left(
t,r,\theta \right) $ are the perturbation functions.

Taking into account only linear on $\delta $ terms, using dimensionless
quantities and supposing that $x_{1}=r_{s}$ and

\begin{gather}
\omega \left( t,r,\theta \right) =\omega \left( r,\theta \right) \exp \left(
-i\sigma t\right) ,  \label{eq13} \\
q_{2}\left( t,r,\theta \right) =q_{2}\left( r,\theta \right) \exp \left(
-i\sigma t\right) ,  \notag \\
q_{3}\left( t,r,\theta \right) =q_{3}\left( r,\theta \right) \exp \left(
-i\sigma t\right) ,  \notag
\end{gather}

\noindent we can obtain from (\ref{eq7}) and second field equation (\ref{eq1}%
)

\begin{gather}
\frac{3\cos \theta \ q_{3}\left( y,\theta \right) }{\sin \theta }%
-7q_{2}\left( y,\theta \right) +8q_{2}\left( y,\theta \right) \ y+2y\left(
y-1\right) \frac{\partial q_{2}\left( y,\theta \right) }{\partial y}+\frac{%
\partial q_{3}\left( y,\theta \right) }{\partial \theta }+  \label{eq14} \\
\frac{2\sigma i}{\epsilon }y^{2}\omega \left( y,\theta \right) =0.  \notag
\end{gather}

\noindent Eq. (\ref{eq14}) requires $\omega \left( y,\theta \right) =0$ if $%
q_{2}\left( y,\theta \right) $, $q_{3}\left( y,\theta \right) $ are real and
$\sigma $ is arbitrary and real. The latter requirement corresponds to the
stability of the perturbed metric (\ref{eq7},\ref{eq12}) in the framework of
the linear stability analysis. More general consideration will be given
elsewhere.

Now we consider first field Eq. (\ref{eq1}) for the perturbed metric. The
obtained expressions are too awkward to be written here and can be found in
\cite{kalashnikov3}. The axial perturbations are defined by the perturbed $%
\left( 1,3\right) $ and $\left( 2,3\right) $ - components of the field
equations. Introducing the new function

\begin{equation}
Q\left( y,\theta \right) =\left[ \frac{\partial q_{3}\left( y,\theta \right)
}{\partial y}-\frac{\partial q_{2}\left( y,\theta \right) }{\partial \theta }%
\right] \sin ^{3}\theta ,  \label{eq15}
\end{equation}

\noindent we can obtain after some manipulations (see \cite{kalashnikov3})
the equations defining the perturbation functions:

\begin{gather}
\frac{\frac{\partial }{\partial \theta }Q\left( y,\theta \right) }{\sin
^{3}\theta }=q_{2}\left( y,\theta \right) \left[ 2-\frac{6}{y}+\frac{%
2y^{2}\sigma ^{2}}{\epsilon }-4\epsilon y^{2}\Delta \left( y\right) \right] ,
\label{eq16} \\
\frac{\frac{\partial }{\partial y}\left( \Delta \left( y\right) Q\left(
y,\theta \right) ^{2}y^{2}\right) }{Q\left( y,\theta \right) \ y\sin
^{3}\theta }=2q_{3}\left( y,\theta \right) \ y^{3}\left[ \frac{\sigma ^{2}}{%
\epsilon }-\epsilon \sin ^{2}\theta \right] ,  \notag
\end{gather}

\noindent where $\Delta \left( y\right) \equiv y^{2}\left( 1-\frac{1}{y}%
\right) $. Now it is possible to eliminate $q_{2}$ and $q_{3}$. This results
in

\begin{equation}
\frac{\partial }{\partial y}\left[ \frac{\frac{\partial }{\partial y}\left(
\Delta \left( y\right) Q\left( y,\theta \right) ^{2}y^{2}\right) }{Q\left(
y,\theta \right) \ y\ \sin ^{3}\theta \ B\left( y\right) }\right] +\frac{%
\partial }{\partial \theta }\left[ \frac{\frac{\partial }{\partial \theta }%
Q\left( y,\theta \right) }{\sin ^{3}\theta \ A\left( y\right) }\right] =%
\frac{Q\left( y,\theta \right) }{\sin ^{3}\theta }.  \label{eq17}
\end{equation}

\noindent Here $A\left( y\right) =-2+\frac{6}{y}-\frac{2y^{2}\sigma ^{2}}{%
\epsilon }$, $B\left( y\right) =$ $-\frac{2y^{3}\sigma ^{2}}{\epsilon }$ and
we used the smallness of $\epsilon $ in order to eliminate the
angular-dependence from $A$ and $B$.

Eq. (\ref{eq17}) permits the separation of variables: $Q\left( y,\theta
\right) \equiv \Xi \left( y\right) \Psi \left( \theta \right) $. Then for
the angular part we have the Gegenbauer's equation:

\begin{equation}
\frac{d^{2}}{d\theta ^{2}}\Psi \left( \theta \right) -3\frac{\cos \theta }{%
\sin \theta }\frac{d}{d\theta }\Psi \left( \theta \right) +2n\Psi \left(
\theta \right) =0,  \label{eq18}
\end{equation}

\noindent where $n$ is some number. Eq. (\ref{eq18}) can be solved through
Legendre functions and then

\begin{equation}
n=\frac{\left( l+2\right) \left( l-1\right) }{2},  \label{eq19}
\end{equation}

\noindent $l$ is integer.

Taking into account that $\epsilon \ll 1$, the equation for the radial part
is

\begin{gather}
2\epsilon y\Delta \left( y\right) \frac{d^{2}}{dy^{2}}\Xi \left( y\right)
+\epsilon y\left( 4y-1\right) \frac{d}{dy}\Xi \left( y\right) +  \label{eq20}
\\
2\Xi \left( y\right) \left[ \sigma ^{2}y^{3}-\epsilon yn-2\epsilon
y+3\epsilon \right] =0.  \notag
\end{gather}

In immediate proximity to the Schwarzschild sphere and outside the $\Xi
\left( y\right) $-extremum (which appears as a result of the angular number
growth, see below Eq. (\ref{eq21})), we can neglect the first term in Eq. (%
\ref{eq20}). Hence, the approximate solution of Eq. (\ref{eq20}) is:

\begin{equation}
\Xi \left( \lambda \right) \approx C\left( 1+\epsilon \lambda \right)
^{6}\left( 3+4\epsilon \lambda \right) ^{\left( -\sigma ^{2}+16\epsilon
\left( n-10\right) \right) \diagup 32\epsilon }\exp \left[ -\frac{\sigma
^{2}\left( 1+\epsilon \lambda \right) \left( 3+2\epsilon \lambda \right) }{%
8\epsilon }\right] ,  \label{eq21}
\end{equation}

\noindent where $C$ is the constant of integration and $\lambda >0$: $%
y\equiv 1+\epsilon \lambda $.

Eq. (\ref{eq21}) suggests that the appropriate asymptotic for $\Xi $ is
provided only by $\sigma ^{2}>0$. Thus, the perturbation increment $\sigma $
is real in the agreement with the stability requirement. The stability far
from $x_{1}$, where the metric coincides with the Schwarzschild one, is
provided by the stability of the latter.

\section{Conclusion}

The numerical simulations based on the Logunov field equations demonstrate a
high accuracy of the approximate analytical solution for the static
spherically symmetric metric in the RTG. It is shown, that there exists a
strong repulsion induced by the massive graviton at the Planckian distance.
As a result of this repulsion, the collapse of the stellar objects has to
entail the burst of the gravitational radiation. However, the rigorous
analysis of the collapse requires taking into account the quantum effects.
Some case for the stability of the collapsar is provided by the stability of
the static spherically symmetric metric.

As the outlooks for the future analysis we can point out the following
topics: i) more general stability analysis including higher-order
perturbations; ii) classical consideration of the possible static
configurations of the matter in the static spherically symmetric metric;
iii) realistic collapse models taking into account the energy loss due to
gravitational radiation in the vicinity of the Schwarzschild sphere.

\section*{Acknowledgments}

Author is Lise Meitner Fellow at Technical University of Vienna and
appreciates the support from the Austrian Science Fund (FWF, grant \#M688).

\end{document}